\documentstyle[prl,aps,epsf]{revtex}

\begin{document}
\draft
\preprint{HEP/123-qed}

\twocolumn[\hsize\textwidth\columnwidth\hsize\csname @twocolumnfalse\endcsname

\title{Direct observation of the washboard noise of a driven vortex lattice in a high-temperature superconductor, Bi$_{2}$Sr$_{2}$CaCu$_{2}$O$_{y}$}
\author{Yoshihiko Togawa$^{1}$, Ryuichi Abiru$^{1}$, Katsuya Iwaya$^{1}$, Haruhisa Kitano$^{1}$, and Atsutaka Maeda$^{1,2}$}
\address{$^{1}$Department of Basic Science, The Univesity of Tokyo, 3-8-1, Komaba, Meguro-ku, Tokyo, 153-8902, Japan\\
$^{2}$CREST, Japan Science and Technology Corporation (JST), 4-1-8, Honcho, Kawaguchi, 332-0012, Japan}
\date{December 30, 1999}
\maketitle
\begin{abstract}
\\
We studied the conduction noise spectrum in the vortex state of a high-temperature superconductor, Bi$_{2}$Sr$_{2}$CaCu$_{2}$O$_{y}$, subject to a uniform driving force. Two characteristic features, a broadband noise (BBN) and a narrow-band noise (NBN), were observed in the vortex-solid phase. The origin of the large BBN was determined to be plastic motion of the vortices, whereas the NBN was found to originate from the washboard modulation of the translational velocity of the driven vortices. We believe this to be the first observation of washboard noise of $dc$ $driven$ vortices in any superconductor. \\
\end{abstract}

\vskip1pc]

\narrowtext

Studies of vortices in high-temperature superconductors (HTSCs) have changed our conventional understanding of vortex matter. One of the most remarkable aspects is the existence of the first-order vortex lattice phase transition (FOT), which separates the vortex liquid phase from the vortex lattice (VL) phase~\cite{Blatter,FOT,Tsuboi1,Fendrich}.

Nonequilibrium states of driven vortices are also of importance. Vortices in motion are characterized by positional and temporal correlations. Theoretical studies of driven vortices in the presence of random pinning~\cite{Koshelev,Balents1,Giamarchi,Balents_vs_Giamarchi,Moon} predicted new dynamic vortex phases, such as a moving-Bragg-glass phase with a power-law decay of positional correlations and no topological defects, and a smectic-flow phase with positional correlations transverse to the flow and none along the flow. In particular, temporal correlations along the flow are considered to be developed only in the moving-Bragg-glass phase. Such phases were observed in the results of numerical simulations~\cite{Moon,Dominguez,Ryu,Olson}. Experimentally, in a conventional type-II superconductor, NbSe$_{2}$, imaging techniques~\cite{Yaron,Matsuda,Troyanovski,Pardo} have showed how vortices move through the sample when they change their configurations. On the other hand, for HTSCs, it is expected that the presence of the FOT can seriously influence the dynamic phases, suggesting new aspects of the vortex dynamics. However, the nature of moving vortices in HTSCs has not been established at all. In addition, comparative studies of driven vortices and many other driven systems, such as density waves and Wigner crystals, should provide important common concepts shared among driven systems with many degrees of freedom. 

The reordering of moving vortices is one of the most fascinating features of driven vortices predicted theoretically~\cite{Koshelev,Balents1,Giamarchi,Balents_vs_Giamarchi,Moon,Dominguez,Ryu,Olson}. The reordered moving medium, which has well-developed temporal correlations, should exhibit a periodic velocity modulation at the so-called washboard frequency~\cite{Balents1}. This effect is well-known for both charge-density waves (CDW)~\cite{Gruner1} and spin-density waves~\cite{Gruner2} systems. Surprisingly, in a recent experiment using scanning tunneling microscopy~\cite{Troyanovski}, the washboard modulation was observed in the creep regime of the VL system of a conventional type-II superconductor, NbSe$_{2}$. In the absence of an external driving force, the coherent motion of just 22 vortices at an extremely slow average velocity ($\sim$nm/s) was observed~\cite{Troyanovski}. Thus, this result was microscopic and it is still basically unknown whether or not the macroscopic moving VL exhibits the washboard velocity modulation. Furthermore, it is important to examine the dynamics of the VL under a controlled external driving force. So far, two ``interference'' experiments have shown the existence of the washboard frequency of the driven VL - one for an Al thin-film~\cite{Fiory} and the other for a YBa$_2$Cu$_3$O$_y$~\cite{Harris}. In these interference experiments, however, the presence of a large ac driving force, which was comparable to the dc driving force, considerably enhanced the coherence of the VL. As a result, the nature of the driven VL was significantly altered. Therefore, direct observation of the washboard modulation in the absence of an external ac current has been eagerly sought.

In this paper, we report the observation of the washboard noise in the conduction noise spectrum under a uniform driving force in Bi$_{2}$Sr$_{2}$CaCu$_{2}$O$_{y}$ (Bi2212). We believe this to be the first observation of the washboard velocity modulation of a $dc$ $driven$ VL in any superconductor. The washboard modulation is characteristic of coherent flow within driven systems. Thus, our result provides a direct observation of the reordering of driven vortices on a macroscopic scale. Our washboard noise data also provide information on the $resistivity$ behavior as a function of magnetic field down to 10$^{-12}$ $\Omega \rm{cm}$. We also found large broadband noise prior to the appearance of the washboard noise, the former being considered to originate from plastic flow of the driven vortices. 

Bi2212 single crystals were grown by the floating-zone method and annealed to an optimally doped state~\cite{Tsuboi1}. They were cut into rectangular pieces with typical dimensions of 1.5$\times$0.5 mm$^{2}$ in the $ab$-plane and 0.015 mm thickness along the $c$-axis. Four-probe dc resistivity measurements showed no indication of other phases, and the resistivity just above the critical temperature $T_{\rm c}$ was $\sim$ 400 $\mu \Omega{\rm cm}$. We will present representative data for a particular piece of crystal with $T_{\rm c}$ of 92.2 K, defined as the zero resistivity temperature. The conduction noise spectra~\cite{Clem} were taken around the FOT in a swept magnetic field with various bias currents. The magnetic field was applied perpendicular to the $ab$-plane, while the current flowed within the plane. The fluctuating voltage between the electrodes was preamplified by a factor of 100 using an SR554 preamplifier and then fed into an HP-35670A FFT analyzer, which can study the noise spectrum with a resolution of 1600 points. The data from each spectrum were typically averaged 100 times. The background noise level was as low as 10$^{-18}$ V$^{2}$/Hz.

Figure\ \ref{fig1} shows the magnetic-field dependence of the resistivity at 80 K, together with the magnetization measured by a SQUID magnetometer. An anomalous change of the magnetization in the reversible region was observed at 70 Oe at 80 K. As has been well established~\cite{Blatter,FOT,Tsuboi1,Fendrich}, this anomaly is associated with the FOT. The resistivity increased rapidly around the FOT, and the field where the rapid increase of the resistivity took place depended on the driving current. However, simultaneous measurements of the magnetization and the resistivity have already clarified that the FOT itself is not influenced by a bias current~\cite{Tsuboi1,Fendrich}. This means that the identification of the FOT using resistivity data for Bi2212 is not good.

Figure\ \ref{fig2} shows the conduction noise spectra for various fields under a bias current of 133 A/cm$^{2}$ at 80 K. We observed two characteristic noise features in the spectra; one is a noise signature which decreases with increasing frequency without any maximum at a finite frequency, and the other is a noise signature which has a peak at a special frequency with a finite bandwidth. In the discussion which follows, we will call the former a broadband noise feature (BBN) and the latter a narrow-band noise feature (NBN). Typical noise spectra are shown in Fig.\ \ref{fig2}. At low magnetic fields, the noise spectra had no characteristic features, as can be seen at 14.4 Oe. As the field increased, the BBN appeared. The detailed behavior of the BBN at 10 Hz as a function of magnetic field is shown in Fig.\ \ref{fig3}. It was clear that the BBN appeared in the field region from 20 Oe to 60 Oe. From 20 Oe to 40 Oe, the intensity of the BBN was so small that it was hard to see the difference in Fig.\ \ref{fig2} (a). However, with increasing field above 40 Oe, the BBN suddenly became approximately one order of magnitude larger. This large BBN decreased rapidly after reaching a maximum at 47.8 Oe. The BBN completely disappeared at around 60 Oe. Once the BBN started to decrease, the NBN appeared. The characteristic feature of the NBN was a rapid shift of the peak frequency to higher frequencies with increasing field. Another interesting point is that the width of the NBN broadened and the height became smaller as the field increased. By increasing the upper limit of the frequency range of the FFT analyzer, the NBN was distinguished from the background noise up to 2900 Hz at 62.2 Oe. Finally, the spectrum became featureless again as shown at the highest field (63.8 Oe) in Fig.\ \ref{fig2} (b). It should be noted that all of these characteristic features in the conduction noise were observed below the FOT field (70 Oe), i.e., in the vortex-solid phase. 

We will first discuss the origin of the large BBN in the field region from 40 Oe to 50 Oe (shown in Fig.\ \ref{fig3}). A lot of similarities were found between the present BBN observed in the conduction noise, and that observed in the local-density noise~\cite{Tsuboi2,Maeda1}. That is, the large BBN appeared in the vortex solid state, reached a maximum just before the resistivity onset and decreased rapidly with further increasing field. The following aspects have already been clarified for the local-density noise~\cite{Tsuboi2,Maeda1}. (1) The presence of both the finger print effect and small spatial correlations showed that the BBN has a bulk origin. (2) Dependence of the spatial correlations on the direction of the Hall probe array suggested a channel-flow-like character. All of these results are consistent with a plastic nature of the driven vortices. In addition, a numerical simulation of the conduction noise~\cite{Olson} showed quite similar behavior, including the large BBN just below the resistivity onset, and the channel-flow-like motion of vortices, both suggesting plastic flow of vortices. Both the local-density noise experiments and the numerical simulation of the conduction noise suggest that the large BBN just below the resistivity onset shown in Fig.\ \ref{fig3} is due to plastic flow of vortices. 

Next, let us discuss the NBN. Periodic modulation in the conduction noise voltage suggests the periodic modulation of the translational velocity of the VL. As discussed above, the most likely candidate for the origin of the periodic velocity modulation is the washboard effect. Simply put, translation of a periodic structure like the VL in the presence of a random pinning force produces an ac component in the velocity at the washboard frequency, $f_{\rm w} = \langle v \rangle / a$. Here $\langle v \rangle$ and $a$ are the averaged velocity and the spacing of the periodic lattice, respectively. Figure\ \ref{fig4} shows the frequency of the observed NBN, $f_{\rm NBN}$, defined as the peak position, as a function of magnetic field. In the same figure, we also plot the expected washboard frequency, $f_{\rm w}$, estimated from the dc resistivity data as follows. $f_{\rm w} = \langle v \rangle / a = (\sqrt{3}/2)^{\frac{1}{2}} \rho j /(B \Phi_{0})^{\frac{1}{2}}$, where $\rho$, $j$, $B$, $\Phi_{0}$ are the resistivity, current density, magnetic field and flux quantum, respectively. In order to estimate $f_{\rm w}$, we used the resistivity data only above 10$^{-9}$ $\Omega \rm{cm}$. Experimentally, the NBN was mostly observed in the field region where the resistivity was below our measurement sensitivity. This means that our NBN data were observed in the so-called creep regime. Thus, in our figure, we have added empirical formulae as guides for the eye, $f \propto \exp (-(H_{\rm 0}/H))$, where $H_{\rm 0}$ is a parameter field. These formulae match the general behavior of the resistivity in the creep regime~\cite{Giamarchi2}. The $f_{\rm NBN}$ data connected with the $f_{\rm w}$ data quite naturally and smoothly, especially at lower current densities (66.7, 133 A/cm$^{2}$). Slight deviations of the resistivity data from the guidelines at higher current densities (267, 400 A/cm$^{2}$) are probably due to the effects of Joule heating. Therefore, our finding of a quite natural and smooth connection between $f_{\rm NBN}$ and $f_{\rm w}$ is strong proof that the NBN observed in the conduction noise resulted from the velocity modulation of the VL at the washboard frequency. 

We believe this to be the first observation of the washboard noise in any superconductor. Since the presence of the washboard noise is a strong indication of the coherent flow of driven vortices, our finding is important in showing how the VL changes from a pinned vortex solid to a coherently moving lattice, via plastic flow, with increasing driving force. We also stress that the data in Fig\ \ref{fig4} reflect for the first time the $resistivity$ behavior (i.e., the velocity of vortices) below the sensitivity of the usual dc resistivity measurement in Bi2212. The magnitude of the resistivity corresponding to the NBN at the lowest magnetic field for 400 A/cm$^{2}$ was 10$^{-12}$ $\Omega \rm{cm}$. 

Some theoretical studies~\cite{Balents1,Moon} have suggested the absence of translational order in the longitudinal direction of the flow and the presence of phase slips between the longitudinal boundaries of elastic domains. As a consequence, the smectic-flow phase becomes preferable to the moving-Bragg-glass phase. These theories have predicted the absence of the washboard noise in driven vortex system. Our observation of the washboard noise directly supports the notion of the development of translational order along the flow direction, and should now force a reconsideration of these theories. The existence of translational order is one of the most important aspects of the moving-Bragg-glass~\cite{Giamarchi}. So, our result may suggest the presence of the moving-Bragg-glass in a HTSC, Bi2212, as well as in a conventional superconductor, NbSe$_{2}$~\cite{Pardo}. 

At first sight, our observation of the BBN before the resistivity onset together with that of the washboard noise in a slightly higher field region are quite consistent with the picture obtained from a numerical simulation~\cite{Olson}. According to Ref.~\onlinecite{Olson}, the pinned VL state changes into coherent flow with washboard velocity modulation, via plastic flow with BBN, as the driving force increases. There are, however, some features inconsistent with the theoretical predictions. The data in Fig.\ \ref{fig2} show that the width of the washboard-noise peak became broader and the height of the peak decreased with increasing magnetic field. These results suggest that the coherence of the VL deteriorates at a higher driving force. The similar dephasing effect of the NBN was also observed in another periodic system, the CDW~\cite{Maeda2}. On the other hand, a numerical simulation based on the Fukuyama-Lee-Rice model for the CDW in quasi-1D chains~\cite{Matsukawa} showed that the width of the NBN becomes sharper and the intensity increases at a higher driving electric field. The theories considering the VL in superconductors~\cite{Koshelev,Balents1,Giamarchi,Balents_vs_Giamarchi,Olson} did not predict any sign of a dephasing effect. Our experimental finding of the dephasing effect might be addressing the very basic questions that are common to the dynamics of quantum condensates with many degrees of freedom. Further work will be needed to clarify these issues. 

In conclusion, we have observed the large BBN and the washboard noise in the conduction noise spectrum of Bi2212 below the FOT field. The large BBN is considered to originate from the plastic motion of driven vortices, while the washboard noise is strong evidence of the coherent nature of moving vortices. We believe this to be the first observation of the washboard noise in any superconductor. This experiment has also provided for the first time the $resistivity$ behavior (i.e., the velocity of vortices) in the creep regime below the sensitivity of usual dc resistivity measurements in Bi2212. Observation of the washboard noise in addition to the BBN is an important first step in exploring the unique nature of the dynamics of the driven VL in HTSCs. 

We thank T. Tsuboi, T. Hanaguri and H. Matsukawa for fruitful discussions and D. G. Steel for a critical reading of the manuscript. This work was, in part, supported by Grant-in-Aid for Scientific Research on Priority Area ``Vortex Electronics". Y.T. thanks the Japan Society for the Promotion of Science for financial support.

\begin{figure}
\caption{Magnetic-field dependence of the dc resistivity under various driving currents (open symbols), together with the magnetization (closed circles), at 80 K. Circles, diamonds, squares and triangles correspond to the resistivity at 66.7, 133, 267 and 400 A/cm$^{2}$, respectively. A solid arrow shows an anomaly of the magnetization associated with the FOT. }
\label{fig1}
\end{figure}

\begin{figure}
\caption{Conduction noise spectra for selected magnetic fields under 133 A/cm$^{2}$ at 80 K. Ticks in the ordinate indicate a noise level of 10$^{-18}$ V$^{2}$/Hz for the data at each magnetic field and solid arrows indicate one decade of the noise power spectral density. Typical BBN and NBN structures are shown in (a) and (b), respectively. The dashed arrows indicate the peak position of the NBN.}
\label{fig2}
\end{figure}

\begin{figure}
\caption{The magnetic-field dependence of the BBN intensity at 10 Hz with 1 Hz band width (closed circles) and the dc resistivity (open circles) under 133 A/cm$^{2}$ at 80 K. }
\label{fig3}
\end{figure}

\begin{figure}
\caption{The NBN frequency $f_{\rm NBN}$ and the expected washboard frequency $f_{\rm w}$. The empirical formulae, $f$ $\propto$ $\exp(-(H_{\rm 0}/H))$, where $H_{\rm 0}$ is a parameter field, are also plotted as guides for the eye. The values of $H_{\rm 0}$ are 2000, 920, 480 and 300 Oe for 66.7, 133, 267 and 400 A/cm$^{2}$, respectively.}
\label{fig4}
\end{figure}

\begin{figure}
\vspace*{1cm}
\leavevmode
\centerline{
\epsfxsize=2.7in
\epsfbox{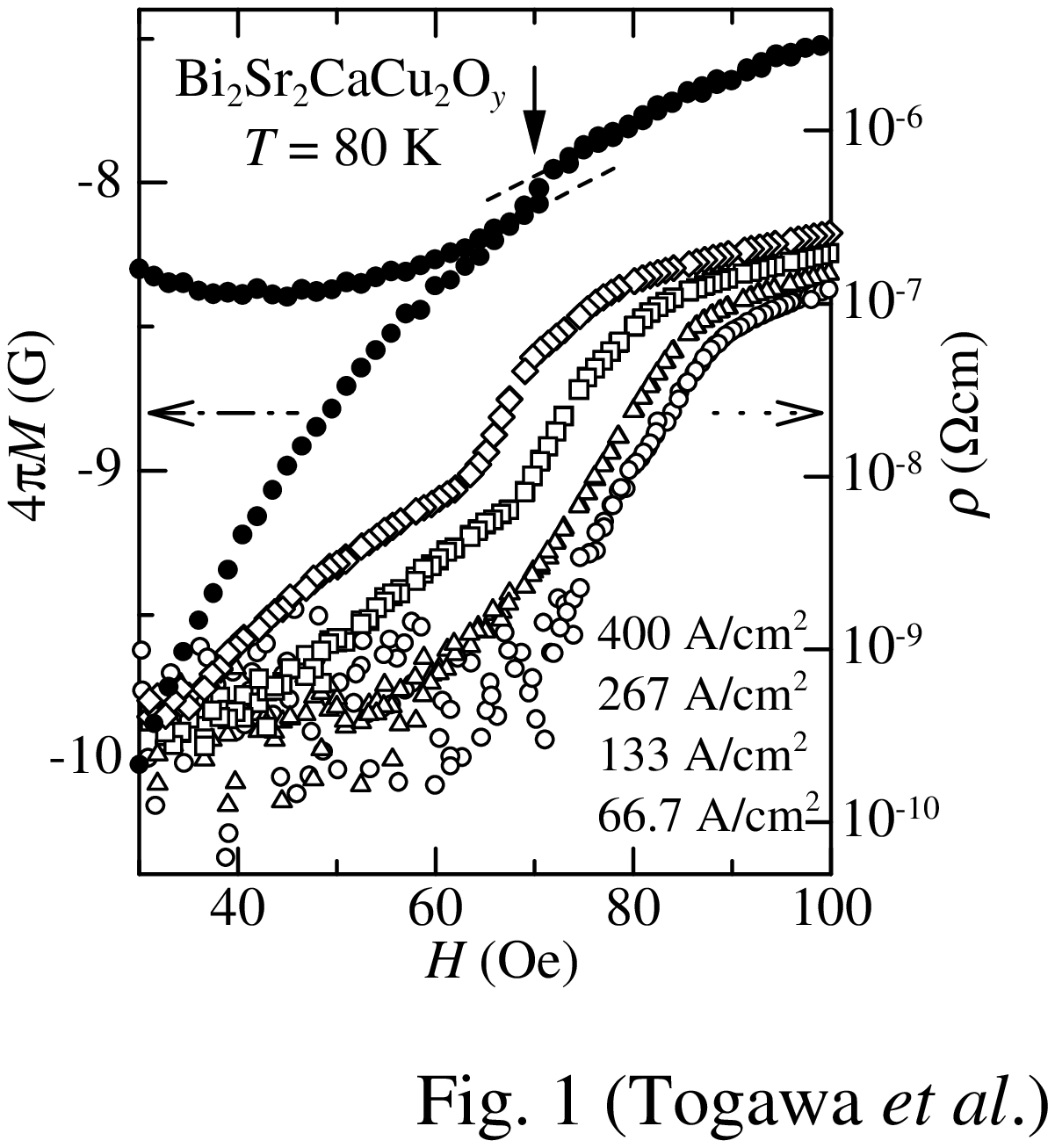}}
\end{figure}

\begin{figure}
\vspace*{3cm}
\leavevmode
\centerline{
\epsfxsize=3.2in
\epsfbox{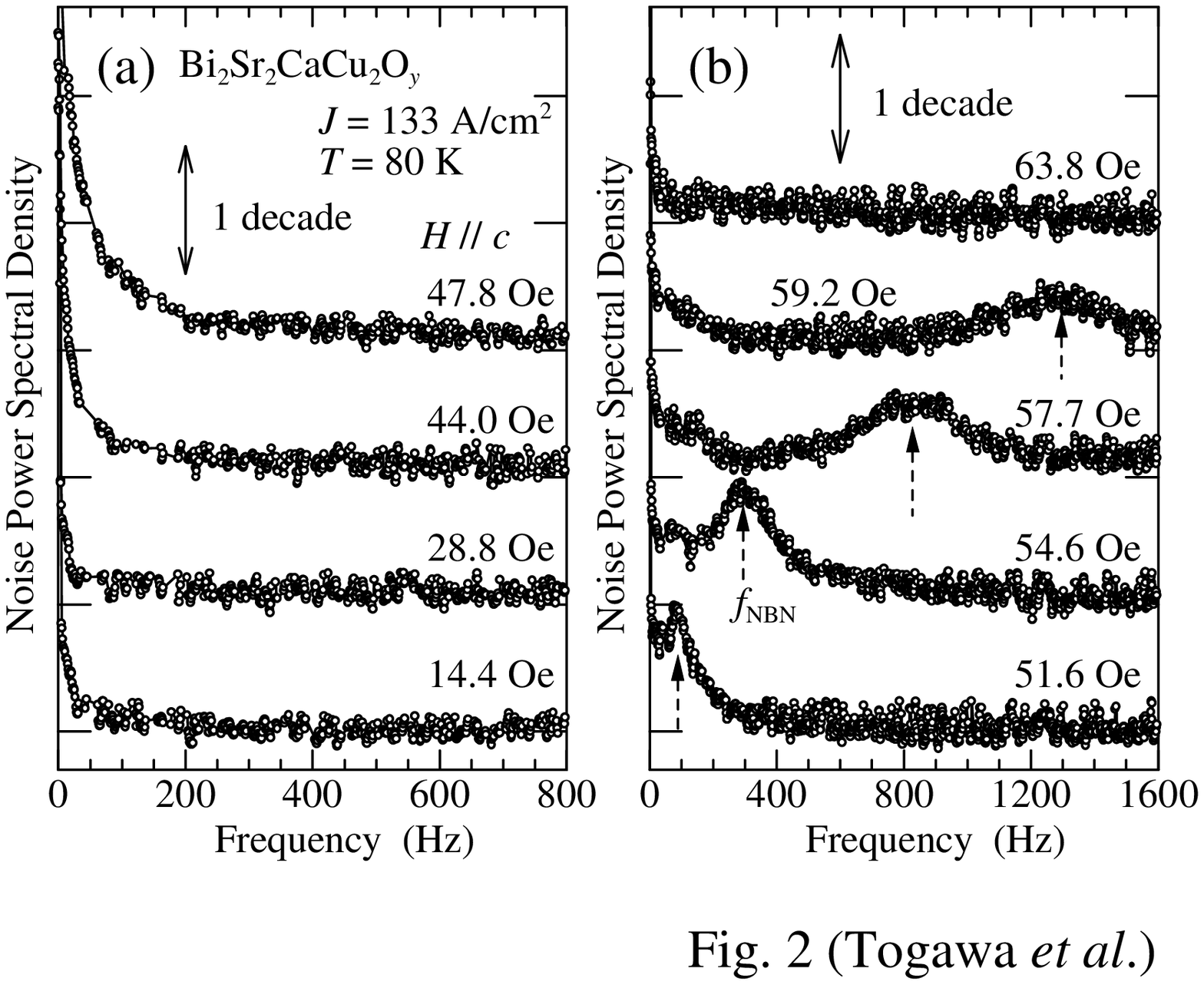}}
\end{figure}

\newpage
\begin{figure}
\vspace*{1cm}
\leavevmode
\centerline{
\epsfxsize=3in
\epsfbox{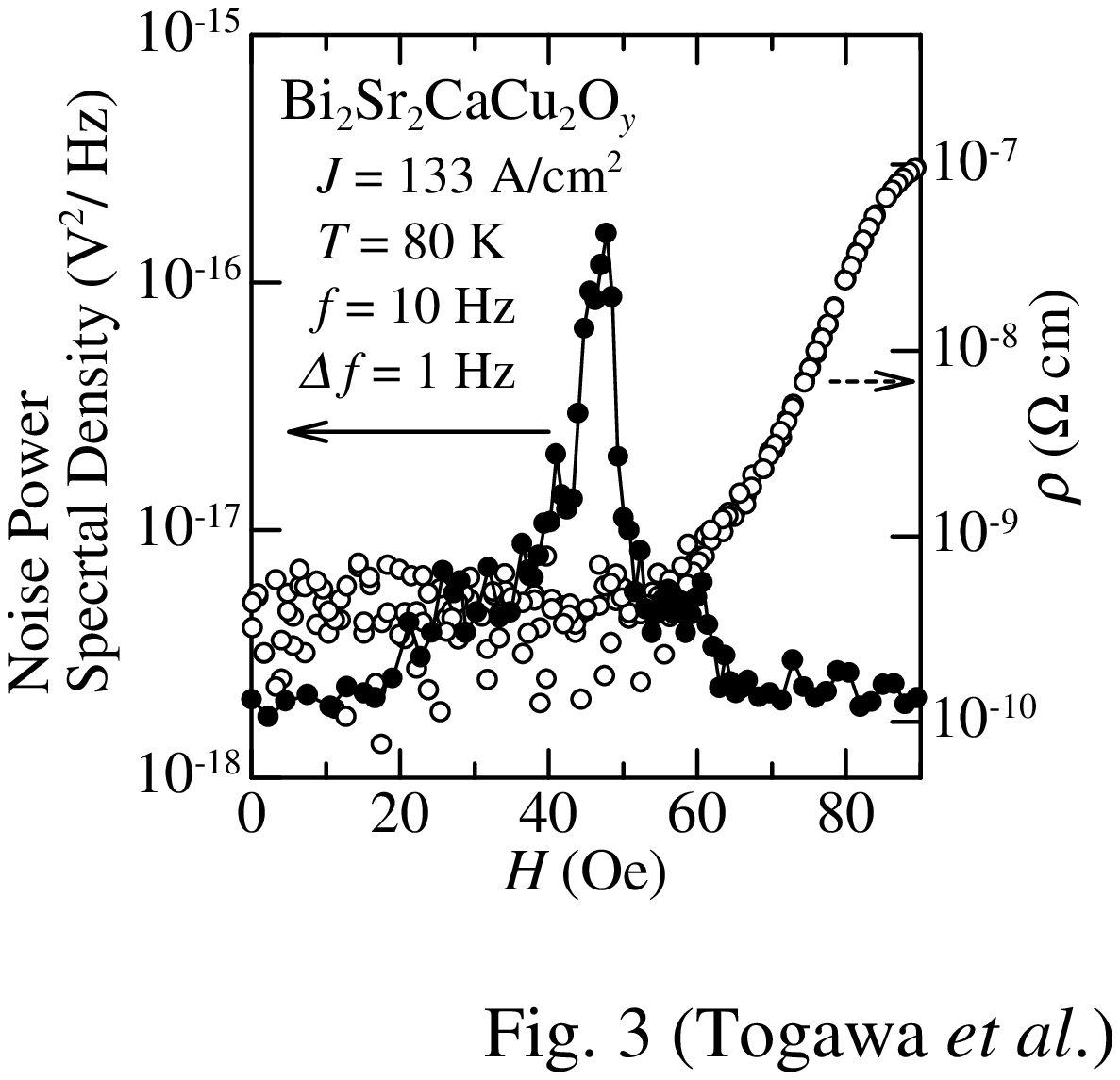}}
\end{figure}

\begin{figure}
\vspace*{3cm}
\leavevmode
\centerline{
\epsfxsize=3in
\epsfbox{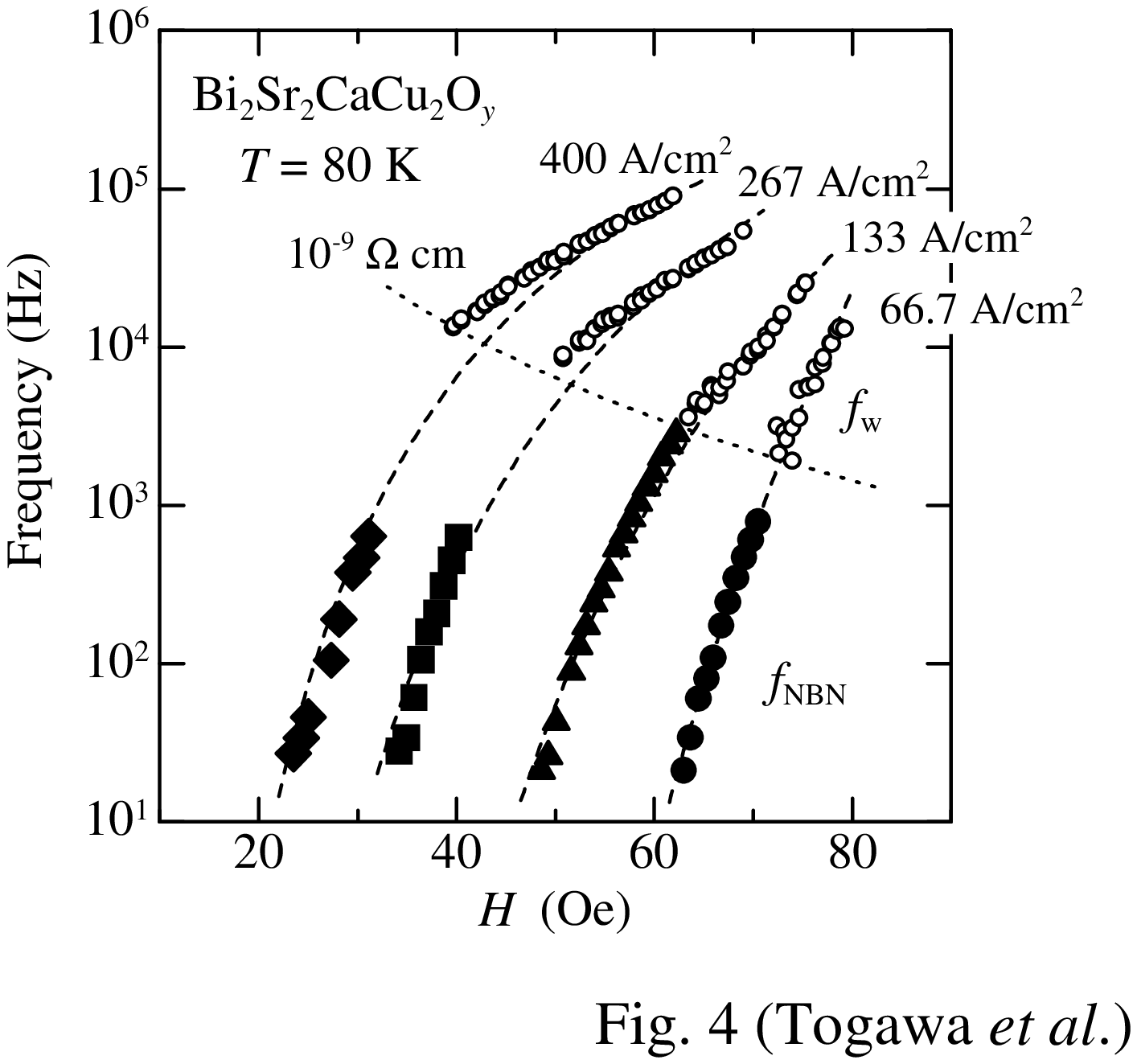}}
\end{figure}

\end{document}